\documentclass[letter, scriptaddress, twocolumn,  prd, showpacs, showkeys]{revtex4-1}

	\usepackage{amsmath}
    \usepackage{mathrsfs}
	\usepackage{amssymb}
	\usepackage{graphicx} 
	\usepackage{makeidx}
	\usepackage{amsfonts}
	\usepackage[ansinew]{inputenc}
	\usepackage[usenames, dvipsnames]{pstricks}
	\usepackage{epsfig}
	\usepackage{pst-grad} 
	\usepackage{pst-plot} 
   \usepackage{todonotes}
	\usepackage[colorlinks, hyperindex]{hyperref}
	\usepackage{float}
	\usepackage{lipsum}
	\usepackage{makecell}
	\usepackage[caption=false]{subfig}
\begin{document} 

\title{The arbitrariness of potentials in interacting quintessence models}
\author{Nandan Roy}  
 \email{royn@nu.ac.th}
 \affiliation{%
The Institute for Fundamental Study ``The Tah Poe Academia Institute", Naresuan University, Phitsanulok 65000, Thailand}

\author{Kazuharu Bamba}%
 \email{ bamba@sss.fukushima-u.ac.jp}
 \affiliation{1Division of Human Support System, Faculty of Symbiotic Systems Science, 
Fukushima University, Fukushima 960-1296, Japan}

\date{\today}

\begin{abstract}

We study the interacting quintessence model with two different types of interaction by introducing a general parameterization of the quintessence potentials. The form of the quintessence potentials is arbitrary as the recent cosmological observations failed to constrain any particular form of the potentials. We explore the interacting quintessence models and investigate if an introduction of interaction between the dark sectors can constrain any particular form of the potential. Our findings reconfirm the arbitrariness of the quintessence potentials even for the interacting dark energy models. As a result, it is shown that the current observations are able to put an upper bound to the interaction parameter for both of the interactions we consider, although it is not possible to constrain the form of the potentials.

\end{abstract}



\maketitle
\section{Introduction} Dark energy is probably the biggest unsolved mystery in  modern cosmology. Many cosmological observations  have confirmed its existence 
\cite{perlmutter1999astrophys, riess1998observational,spergel2007three,tegmark2004cosmological,seljak2005cosmological,ade2016planck}  but the exact nature of it is still unknown. The Lambda Cold Dark Matter Model ($\lambda$CDM) \cite{Weinberg:1988cp, Peebles:2002gy,padmanabhan2003cosmological, Sahni:1999qe,Sahni:1999gb,Bianchi:2010uw} is the most popular and observationally consistent model of dark energy in which the cosmological constant is considered as the candidate of the dark energy.


Although $\lambda$CDM is successful and consistent in explaining the accelerated expansion of the universe, it still has to overcome some challenges coming from both observations and theoretical interpretation of the cosmological constant.  Recently a substantially discordance at the level of $2.3 \sigma $ in the ($\lambda$CDM) model is reported between the Planck 2015 CMB data and KiDS(Kilo Degree Survey) data \cite{ ade2016planck, hildebrandt2016kids, de2015first,kuijken2015gravitational, fenech2017calibration}. There are other observational challenges of the $\lambda$CDM model, one of which comes from the $3 \sigma$ tension between the direct measurement Hubble constant from Hubble Space Telescope data (HST) and the value of Hubble constant obtained from the CMB data considering the $\lambda$CDM model \cite{riess20113, riess20162}.  In field theory, the cosmological constant can be interpreted as constant vacuum energy.  The interpretation of the cosmological constant as vacuum energy also give rise to theoretical challenge to the $\lambda$CD model which is known as famous cosmological constant problem. The problem comes from the discrepancy between the theoretical prediction and the observed value of the cosmological constant which is of the order of $ 10^{121}$.

Consideration of dynamical dark energy models is one way to alleviate the cosmological constant problem \cite{copeland2006dynamics}. Another way is the  modification of the theory of gravity, which can also explain the accelerated expansion without considering any exotic matter component, but recently these models are weakened by the estimation of the speed of gravitational wave as the speed of the light from the observation of the two binary systems of neutron stars colliding \cite{Ezquiaga:2017ekz,Sakstein:2017xjx,Creminelli:2017sry,lombriser2016breaking,lombriser2017challenges}. For recent reviews on the issue of dark energy and the modification of gravity theory to explain the late-time cosmic acceleration, see, for instance,~\cite{nojiri2011unified,capozziello2011extended,nojiri2017modified,capozziello2010beyond,bamba2012dark, bamba2015inflationary, cai2016f, nojiri2017modified}

Quintessence dark energy models are the most popular amongst the dynamical dark energy models \cite{ratra1988cosmological,tsujikawa2013quintessence,bamba2012dark}. In these models of dark energy the accelerated expansion is driven by a non minimally coupled scalar field and an associated quintessence potential. The form of the quintessence potential is arbitrary as there is no identification of a particular form of the quintessence potentials either from the cosmological observations or from basic physics. This allowed researchers to consider a wide range of potentials and all of them in certain parameter range satisfy the observations \cite{Copeland:1997et,caldwell1998cosmological,zlatev1999quintessence,de2000cosmological,ng2001applications,corasaniti2003model,linder2006paths}. Recently a general parameterization of the quintessence potentials has been proposed in \cite{roy2018new}. The general parameterization includes three parameters which are named as dynamical parameters or  $\alpha$ parameters. By tuning these $\alpha$ parameters one can get back different quintessence potentials. This general form has been implemented in the Boltzmann code CLASS and MCMC code Montepython to check if there is any preference on the values of the $\alpha $ parameters. It has been shown in the above mentioned work that the recent cosmological observations have failed to constrain the $\alpha $ parameters and which actually means that the arbitrariness of the quintessence potentials can not be resolved by the recent cosmological observations,.

In this work we have extended the analysis to the interacting quintessence models. The same general parameterization of the quintessence potentials has been considered along with two different kind of interactions between the dark sectors and we have tried to find out if there is any change in the finding of \cite{roy2018new}. Like quintessence potentials the functional form of the coupling between the dark sectors is also arbitrary. The form of the interactions we have considered is for example by keeping in mind the mathematical simplicity as our aim is to find if there is any effect of the interaction between the dark sectors on constraining the forms of the potential.

The field equations are written as a set of autonomous equations through a suitable variable transformation.  Later on, these autonomous equations are transferred in to polar form \cite{roy2018new, Roy:2013wqa,Urena-Lopez:2015gur, Urena-Lopez:2015odd,Roy:2014yta, reyes2010attractor}. The particular importance of the polar form of the system is the direct relation of the cosmological variables to the dynamical variables. We have considered two different types of interactions between the dark sectors and used the cosmological observations to constrain the parameters in the model. 

Following is the summary of the paper. In Sec.II, we have discussed about the mathematical background of the system. Sec.III deals with the interactions and the corresponding system of equations. In Sec. IV, we have shown how to estimate a set of initial conditions of the system by matching matter dominated and radiation dominated approximate solutions. Sec.V presents a full Bayesian analysis of the model using diverse cosmological observations to constrain different parameters in the model. In Sec.VI We give a summary and conclusion of the results we obtained.

\section{Mathematical Background}
Let us consider the universe to be filled with radiation, dark matter and the dark energy. The dark energy is in the form of quintessence scalar field. We also consider that there is an interaction between the dark matter and the dark energy. The components of the universe are barotropic in nature, hence they obey the relation $p_j = w_j \rho_j$, for radiation $w_r = 1/3$ and for dark matter $w_m = 0$. In such a universe which is spatially flat, homogeneous and isotropic  the Einstein field equations are written as
\begin{subequations}
\label{eq:field}
  \begin{eqnarray}
    H^2 &=& \frac{\kappa^2}{3} \left( \sum_j \rho_j +
      \rho_\phi \right) \, , \label{eq:field1} \\
    \dot{H} &=& - \frac{\kappa^2}{2} \left[ \sum_j (\rho_j +
      p_j ) + (\rho_\phi + p_\phi) \right] \,,
      \label{eq:field2} 
  \end{eqnarray}
\end{subequations}
where $\kappa^2 = 8 \pi G$ and $a$ is the scale factor of the Universe, and $H \equiv \dot{a}/a$ is the Hubble parameter. A dot means derivative with respect to
cosmic time. The continuity equations of the radiation, matter and the scalar field can be respectively written as 

\begin{subequations}
\begin{eqnarray} \label{eq.conty}
\dot{\rho_r} + 3 H \rho_r (1 + w_r) &=& 0, \\ \label{con_rad}
\dot{\rho_m} + 3 H \rho_m (1 + w_m) &=& - Q , \\ \label{con_mat}
\dot{\rho_\phi} + 3 H (\rho_\phi + p_\phi) &=& + Q. \label{con_phi}
\end{eqnarray}
\end{subequations}
The wave equation of the scalar field is written as
\begin{equation} \label{eq:wave}
\ddot{\phi} + 3 H \dot{\phi} + \frac{dV}{d\phi} = \frac{Q}{\dot \phi},
\end{equation}

where $V$ is the scalar field potential related to the quintessence field and $Q$ is the interaction term between the dark energy and dark matter. A positive coupling $Q$ indicates an exchange of energy from dark matter to dark energy and a negative $Q$ indicates an exchange of energy from dark energy to dark matter.

    We introduce following sets of dimensionless variables to write the systems of equations as an set of autonomous equations,

\begin{subequations}
  \label{eq:3}
  \begin{eqnarray}
    x &\equiv& \frac{\kappa \dot{\phi}}{\sqrt{6} H} \, , \quad y
               \equiv \frac{\kappa V^{1/2}}{\sqrt{3} H} \, , \label{eq:3a}
    \\
    y_1 &\equiv& - 2\sqrt{2} \frac{\partial_{\phi} V^{1/2}}{H} \, , \quad
                 y_2 \equiv - 4\sqrt{3} \frac{\partial^2_\phi
                 V^{1/2}}{\kappa H} \, . \label{eq:3b}             
  \end{eqnarray}
\end{subequations} 

This particular transformation was first used in \cite{Urena-Lopez:2015gur} and later it is used in \cite{roy2018new}. Using these sets of new variables the system of equations which governs the dynamics of the scalar field reduces to the following set of autonomous equations,

\begin{subequations}
  \label{eq:system}
  \begin{eqnarray}
    x^\prime &=& - \frac{3}{2} \left(1 - w_{tot} \right) x + \frac{1}{2} y y_1 + q \,
                 , \label{eq:4a} \\
    y^\prime &=&  \frac{3}{2} \left(1 + w_{tot} \right) y - \frac{1}{2} x y_1 \,
                 , \label{eq:4b} \\
    y^\prime_1 &=&  \frac{3}{2} \left(1 + w_{tot} \right) y_1  + x
                   y_2 \, , \label{eq:4c}
  \end{eqnarray}
\end{subequations}

where $q = \frac{\kappa Q}{\sqrt[]{6} H^2 \dot{\phi}}$ and a `prime' represents differentiation with respect to the e-foldings $N = \ln(a)$ and the present value of the scale factor is scaled to be unity. The $w_{tot}$ is the total equation of state of the system which is defined as $w_{tot} \equiv \frac{p_{tot}}{\rho_{tot}} = \frac{1}{3} \Omega_r + x^2
- y^2$. 

\subsection{Polar form}
Further we introduced a set of polar transformation to write equations(\ref{eq:system}) in a polar form. Following are the polar transformations, $x = \Omega_{\phi} ^{1/2} \sin(\theta/2)$ and $y = \Omega_{\phi}^{1/2}
\cos(\theta/2)$, where $\Omega_{\phi} = \kappa^2 \rho_{\phi}/3 H^2$ and the $\theta$ is angular degree of freedom. With these transformations the equations in (\ref{eq:system}) reduces to

\begin{subequations}
\label{eq:polar}
  \begin{eqnarray}
    \theta^{\prime} &=& - 3 \sin \theta + y_1 + q \ \Omega_{\phi} ^{1/2} \cos(\theta/2)\, , \label{eq:10a} \\
    y_1^{\prime} &=& \frac{3}{2} \left( 1  + w_{tot} \right) y_1
                     + \Omega_{\phi} ^{1/2} \sin(\theta /2) y_2  \,
                     , \label{eq:10b} \\ 
    \Omega_{\phi} ^{\prime} &=& 3 (w_{tot} - w_{\phi})
                                \Omega_{\phi} + q \ \Omega_{\phi} ^{1/2} \sin(\theta/2) \, . \label{eq:10c}       
  \end{eqnarray}
\end{subequations}
 The advantage of writing down the equations(\ref{eq:system}) in polar form is the direct connection between the cosmological variables and the dynamical variables whereas in the previous transformations the $x$ and $y$ does not carry any direct physical meaning. The $\Omega_{\phi}$ is itself the scalar field energy density and the the $\theta $ can be directly related to the equation of state of the scalar field as $w_{\phi} = - \cos \theta $. Apart from this $\theta$ can also give us information about the ratio of K.E and potential energy of the scalar field as $\tan^2 \theta = \frac{\frac{1}{2} \dot{\phi}}{ V(\phi)} = \frac{x^2}{y^2}$. One can see from the equations in  Eq.(\ref{eq:polar}) that the system of equations is not close until one has the information about the potential variable $y_2$ related to the potential $V(\phi)$ and the interaction variable $q$ related to the interaction term $Q$. To close the above system of equations we must consider some particular form of the $V(\phi)$ and $Q$.  
 
 
 Unfortunately both of these two quantities are arbitrary as there is no preferential functional form of $y_2$ and $Q$ from the cosmological observations or from the basic physics. In a very recent work by Roy et al. \cite{roy2018new} a general form of $y_2$ has been proposed which includes almost all popular form of the quintessence potentials. In this work we will consider the same function form of $y_2$ as $\frac{y_2}{y} = \alpha_0 + \alpha_1 (\frac{y_1}{y}) + (\frac{y_1}{y})^2$ together with two different form of the coupling $Q$. The specialty of this general parameterization is that  by choosing different value of the $\alpha_i$ one can select a particular form of the potential. For example in Table 1, we have included the same classifications of the potential form the \cite{roy2018new}. In the same work it has been also shown that for a non interacting quintessence model it is not possible to constrain the $\alpha$ parameters hence there is no preferred choice for the quintessence potential. In this work we plan to check if an interaction between the dark sectors of the universe can put any constrain the $\alpha$ parameters. 

\begin{table*}[htp!]
\caption{\label{tab:1} A classification of different potentials depending on the choice of $\alpha$ parameters given in \cite{roy2018new}.}
\begin{ruledtabular}
\begin{tabular}{|c|c|c|}
No & Structure of $y_2/y$ & Form of the potentials $V(\phi)$  \\ 
\hline 
Ia&  $\alpha_0 = 0, \alpha_1 = 0, \alpha_2 \neq  - \frac{1}{2}$ & $ (A  + B \phi)^{\frac{2}{(2 \alpha_2 +1)}}$ \\
\hline
Ib & $\alpha_0 = 0, \alpha_1 = 0, \alpha_2 = - \frac{1}{2}$ & $A^2 e^{2 B \phi}$ \\
\hline 
IIa & $\alpha_0 \neq 0, \alpha_1 = 0, \alpha_2 \neq - \frac{1}{2}$& $A^2 \cos \left[ \sqrt{\alpha_0 \kappa^2 (1 + 2 \alpha_2)} (\phi -  B) /2 \sqrt{3} \right]^{\frac{2}{1 + 2 \alpha_2}}$ \\
\hline 
IIb & $\alpha_0 \neq 0, \alpha_1 = 0, \alpha_2 = - \frac{1}{2}$ & $A^2 \exp \left({- \kappa^2 \alpha_0  \phi^2/12}) \exp({ 2 B \phi} \right)$ \\
\hline
IIIa & $\alpha_0 = 0, \alpha_1 \neq 0, \alpha_2 \neq - \frac{1}{2}$ & $\left[ A \exp \left( \alpha_1 \kappa \phi/\sqrt{6} \right) + B\right]^{\frac{2}{1+2 \alpha_2}}$ \\
\hline
IIIb & $\alpha_0 = 0, \alpha_1 \neq 0, \alpha_2 = - \frac{1}{2}$ & $A^2 \exp \left[ 2 B ~ \exp \left(	\kappa \alpha_1 \phi /\sqrt{6} \right) \right] $ \\
\hline
IVa & $\alpha_0 \neq 0, \alpha_1 \neq 0, \alpha_2 \neq - \frac{1}{2}$ & $A ^2 \exp(\frac{ \kappa \alpha_1 \phi}{ \sqrt{6} ( 1 + 2 \alpha_2)}) \left\{ \cos \left[ \left(- \frac{\kappa^2 \alpha_1 ^2}{24} + \frac{\kappa^2 \alpha_0}{12} (1 + 2 \alpha_2) \right)^{\frac{1}{2}} (\phi - B) \right] \right\}^{\frac{2}{1+2 \alpha_2 }}$ \\
\hline
IVb & $\alpha_0 \neq 0, \alpha_1 \neq 0, \alpha_2 = - \frac{1}{2}$ & $A^2 \exp \left[ \frac{\kappa \alpha_0 \phi}{\sqrt{6} \alpha_1	} + 2B \exp \left(	\frac{\kappa \alpha_1 \phi}{\sqrt{6}} \right) \right]$ 
\end{tabular}
\end{ruledtabular}
\end{table*}

\section{Interaction}

We have consider two different interaction terms for the analysis. Likewise the quintessence potentials the form of the interaction term is also arbitrary and each one of them can explain accelerated expansion for certain parameter range. For a list of different types of interactions in quintessence models we refer to \cite{Bahamonde:2017ize}. The reason behind our choice of interaction terms is the mathematical simplicity of the field equations particularly to be able to write the interaction term $q$ as a function of $\Omega_\phi$ and $\theta$ so that we can close the systems of equations in Eq.(\ref{eq:polar}).


\subsection{$ Q = \beta H \dot{\phi}^2 $ }
We consider the interaction to be of the form $Q = \beta H \dot{\phi}^2$ \cite{mimoso2006asymptotic}. After considering this particular form of the coupling  the equations in (\ref{eq:polar}) reduces to

\begin{subequations}
\label{eq:coup1}
  \begin{eqnarray}
    \theta^{\prime} &=& - 3 \sin \theta + y_1 + \frac{\beta}{2} \Omega_{\phi}  \sin \theta \, , \label{eq:10a} \\
    y_1^{\prime} &=& \frac{3}{2} \left( 1  + w_{tot} \right) y_1
                     + \Omega_{\phi} ^{1/2} \sin(\theta /2) y_2  \,
                     , \label{eq:10b} \\ 
    \Omega_{\phi} ^{\prime} &=& 3 (w_{tot} - w_{\phi})
                                \Omega_{\phi} + \frac{\beta}{2} \  \Omega_{\phi} \  (1 + w_\phi) \, . \label{eq:10c}       
  \end{eqnarray}
\end{subequations}


\subsection{$Q = \beta (3 H^2 - \kappa \rho_{\phi}) \dot{\phi}^2 / H $}

With the consideration of the above functional form of the interaction the system of equations in Eq.(\ref{eq:polar}) reduces to the following sets of equations,

\begin{subequations}
\label{eq:coup2}
  \begin{eqnarray}
    \theta^{\prime} &=& - 3 \sin \theta + y_1 + \frac{3 \beta}{2} \Omega_{\phi} (1 - \Omega_{\phi}) \sin \theta \, , \label{eq:10a} \\
    y_1^{\prime} &=& \frac{3}{2} \left( 1  + w_{tot} \right) y_1
                     + \Omega_{\phi} ^{1/2} \sin(\theta /2) y_2  \,
                     , \label{eq:10b} \\ 
    \Omega_{\phi} ^{\prime} &=& 3 (w_{tot} - w_{\phi})
                                \Omega_{\phi} + \frac{3 \beta}{2} \  \Omega_{\phi} \ (1 - \Omega_{\phi}) \  (1 + w_\phi) \, . \label{eq:10c}       
  \end{eqnarray}
\end{subequations}

We have modified the publicly available CLASS \cite{lesgourgues2011cosmic} code to incorporate these two sets of equations in Eq.(\ref{eq:coup1}) and Eq.(\ref{eq:coup2}) separately. In the Section V more details about the  modification of the CLASS code is discussed. For the numerical analysis with the class code to be done one has to supply a viable initial condition to the CLASS code. Rather be choosing initial condition arbitrarily in the next section we have follow the method of estimating the initial condition from \cite{roy2018new}. 

\section{Initial conditions }
In this section we will try to estimate the initial conditions of the universe based on the approximate solution of the matter and radiation dominated era and match them at the radiation-matter equality. From the recent cosmological observations the present value of the dark energy equation of state $w_\phi \simeq -1$ and this implies that $\theta < 1$. We have also assume that when the universe entered in the matter dominated phase from the radiation dominated era till then the dark energy density was very sub-dominant $\Omega_\phi \ll 1$. It is important to mention here that this particular method of finding the initial conditions is for the thawing type potentials. Our assumptions $\Omega_\phi \ll 1$ and $y_1 > 0$ at the beginning which means the EOS of dark energy deviates from the cosmological constant at the late time essentially falls in the classification of the thawing potentials. By considering these two fact $\theta \ll 1$ and $\Omega_\phi \ll 1$ the equations in (\ref{eq:coup1}) and (\ref{eq:coup2}) reduces to the following form,

\begin{subequations}
\label{eq:ini}
  \begin{eqnarray}
    \theta^{\prime} &=& - 3 \theta + y_1 \, , \label{eq:ini_a} \\
    y_1^{\prime} &=& \frac{3}{2} \left( 1  + w_{tot} \right) y_1 \,
                     , \label{eq:ini_b} \\ 
    \Omega_{\phi} ^{\prime} &=& 3 (w_{tot} - w_{\phi})
                                \Omega_{\phi} \, . \label{eq:ini_c}       
  \end{eqnarray}
\end{subequations}

Considering the fact that  $\theta \ll 1, \ \sin(\theta) \simeq \sin(\theta/2) \ll 1 $ and $\Omega_\phi \ll 1$ we have neglected the last term from the equations in (\ref{eq:coup1})
 and (\ref{eq:coup2}). The approximate form of the equations in (\ref{eq:ini}) is valid for both forms of the interaction. Now we shall try to find out solution for radiation and matter dominated era separately and later match them at the radiation matter equality which will allow us to make a good guess about the initial condition of the universe that can evolve to give us a present day accelerating universe. The $e$-folding is different for radiation and matter dominated era. For radiation dominated era $N_r = \ln(a/a_i)$ and for matter dominated era $N_m = \ln(a/a_{eq})$ where $a_i$ is the initial value of the scale factor, whereas $a_ {eq} $ is the value of the scale factor at the radiation-matter equality.                  
\subsection{Radiation Dominated Era}
In a radiation dominated universe the total EOS is $w_{tot} = 1/3$ and the equations in (\ref{eq:ini}) reduces to 

\begin{eqnarray}
  \theta^{\prime} = - 3 \theta + y_1 \, , \quad y_1^{\prime} = 2 y_1
  \, , \quad \Omega_{\phi}^{\prime} = 4 \Omega_{\phi}\,
  . \label{eq:rad}       
\end{eqnarray}
Considering the growing solutions in the radiation dominated era the approximate solutions are given by,

\begin{equation}
  \theta_r = \theta_i (a/a_i)^2 \, , \; y_{1r} = y_{1i} (a/a_i)^2 \,
  , \; \Omega_{\phi r} =  \Omega_{\phi i } (a/a_i)^4 \,
  , \label{eq:sol_rad}
\end{equation}
The sub-index `$r$' represents the solutions during the radiation dominated era and `$i$' represents initial value of the cosmological parameters. In addition to these solution we found another relation between $y_1$ and $\theta$ as $y_1 = 5 \theta$.

\subsection{Matter Dominated Era}

Once the universe is matter dominated the total EOS is $w_{tot} \simeq 0$. The equations in (\ref{eq:ini}) reduces to

\begin{equation}
  \theta^{\prime} = - 3 \theta + y_1 \, , \quad y_1^\prime =
  \frac{3}{2} y_1 \, , \quad \Omega_{\phi}^{\prime} = 3 \Omega_{\phi}
  \, . \label{eq:matter}       
\end{equation}

The solution of these equations are given by,

\begin{eqnarray}
  \theta_m &=& \left( \theta_{eq} - \frac{2}{9} y_{1eq} \right)
  (a/a_{eq})^{-3} + \frac{2}{9} y_{1eq} (a/a_{eq})^{3/2} \, ,
  \nonumber \\ 
  y_{1m} &=& y_{1eq} (a/a_{eq})^{3/2} \, , \; \Omega_{\phi m} =
  \Omega_{\phi eq} (a/a_{eq})^3 \, . \label{eq:sol_matt}     
\end{eqnarray}

A sub-index $`m$' represents solutions at the matter dominated era and $`eq$' represents value of the cosmological parameters at the radiation matter equality. We do not neglect the decaying solutions in contrast with the radiation dominated solution as it will make the matching of the solutions at the radiation-matter equality simpler. Once we perform the matching of the equations (\ref{eq:sol_rad}) and (\ref{eq:sol_matt}) at the radiation matter equality $a_{eq} = \Omega_{r0}/\Omega_{m0}$ it allowed us to find a solution at the matter domination which has the information about the initial sate of the universe.  From the matching of the solutions we found  $\theta_{eq} = \theta_i (a_{eq}/a_i)^2$, $y_{1eq} = 5\theta_{eq} = y_{1i} (a_{eq}/a_i)^2$ and $\Omega_{\phi eq} = \Omega_{\phi i} (a_{eq}/a_i)^4$, and substituting it in (\ref{eq:sol_matt}) obtain
\begin{subequations}
\label{eq:matter-dom}
\begin{eqnarray}
\theta_m &=& \frac{10}{9} \left (\frac{a_{eq}}{a_i} \right) ^2\, \theta_i \left[ \left( \frac{a}{a_{eq}} \right)^{3/2} - \frac{1}{10}  \left( \frac{a}{a_{eq}} \right)^{-3} \right] \, , \label{eq:matter-doma} \\
y_{1m} &=& \frac{a^{1/2}_{eq}}{a^2_i} \, y_{1i} \, a^{3/2} \, , \label{eq:matter-domb} \\
\Omega_{\phi m} &=& \frac{a_{eq}}{a_i^4} \, \Omega_{\phi i}  \, a^3\, . \label{eq:matter-domc}
\end{eqnarray}
\end{subequations}

 Hence by considering the present value of the scale factor to be unity or $a = 1$ in (\ref{eq:matter-dom}) we estimate initial condition for the dynamical variables as 

\begin{subequations}
\label{eq:7}
\begin{eqnarray}
  \theta_i &\simeq& \frac{9}{10} a^2_i \,
  \frac{\Omega^{1/2}_{m0}}{\Omega^{1/2}_{r0}} \theta_0
  \, , \label{eq:7a} \\
  \Omega_{\phi i} &\simeq& a_i^4 \, \frac{\Omega_{m0}}{\Omega_{r0}} \, \Omega_{\phi 0} \, , \label{eq:7b}
\end{eqnarray}
\end{subequations}

The initial values of the dynamical variables $\theta$ and $\Omega_{\phi}$ can be estimated from the present values of them and one can estimate $y_{1i}$ from the relation $y_{1i} = 5 \theta_i$.

\section{Numerical Investigation}
 In this section we discuss about the general method we adopt to constrain the dynamical variables and the cosmological parameters using cosmological observations. 

\subsection*{Data sets}

For the numerical investigation of the system we have used an modified version of the Boltzmann code CLASS \cite{lesgourgues2011cosmic} and the MCMC code Monte Python \cite{Audren:2012wb}. Modification to the CLASS code is done separately for two different interactions. In order to maintain the interface with Montepython so that it can sample all extra dynamical parameters which we are having in our analysis, the necessary modifications are also done to the CLASS code. 

At the beginning of any numerical run it is necessary to fine tune the initial values of the dynamical variables. To fine tune the initial conditions we write $y_{1i} = 5\theta_i$, $\theta_i = P \times$Eq.~\eqref{eq:7a} and $\Omega_{\phi i} = Q \times$Eq.~\eqref{eq:7b}. The values of $P$ and $Q$ are adjusted by the shooting method which is already implemented in the CLASS code for the scalar field. Generally the $P,Q = \mathcal{O}(1)$ is enough to find a successful initial condition which can give us $\Omega_{\phi 0}$ and $w_{\phi 0}$ with a very high precision.

 We will be sampling all the $\alpha$ parameters  $\alpha_0, \alpha_1, \alpha_2$ in the general form of the potential together with other $\Omega_\phi, w_\phi, y_1$ and the interaction parameter $\beta$. Sampling of the $\alpha$ parameters will allow us to sample the general form of the potential and sampling  of the interaction parameter $\beta$ will carry information about the energy transfer between dark matter and dark energy.  

We have used three data sets: (i) the SDSS-II/SNLS3 JLA supernova data\cite{Betoule:2014frx}  and (ii) BAO measurements data and (iii)  Hubble data. The back ground quantities are sensitive to these data sets. A Planck2015 prior has been imposed on the baryonic and cold dark matter sector \cite{Adam:2015rua, Aghanim:2015xee, Ade:2015rim,Ade:2015xua,Ade:2015lrj}: $\omega_b = 0.02230 \pm 0.00014 $ and $\omega_{cdm} = 0.1188 \pm 0.0010$. A flat prior has been imposed on the $\alpha$ parameters and the $\beta$ parameter. Following is the prior which we have considered for the data analysis $-20<\alpha_0<20$,  $-5<\alpha_1<5$, $-2<\alpha_2<2$ and $0 < \beta < 30$. We have consider only positive prior for $\beta$ as we consider transfer of energy from dark matter  to dark energy. All these parameters are sampled by the MCMC code Monte Python. The set of derived parameters are $\Omega_m$, $\Omega_{\phi}$, $w_{\phi}$ and $y_1$. In what follows we discuss about the result we obtain from the numerical analysis.


\subsection{$ Q = \beta H \dot{\phi}^2 $ }

\begin{figure}[] 
\centering
\includegraphics[width= \columnwidth]{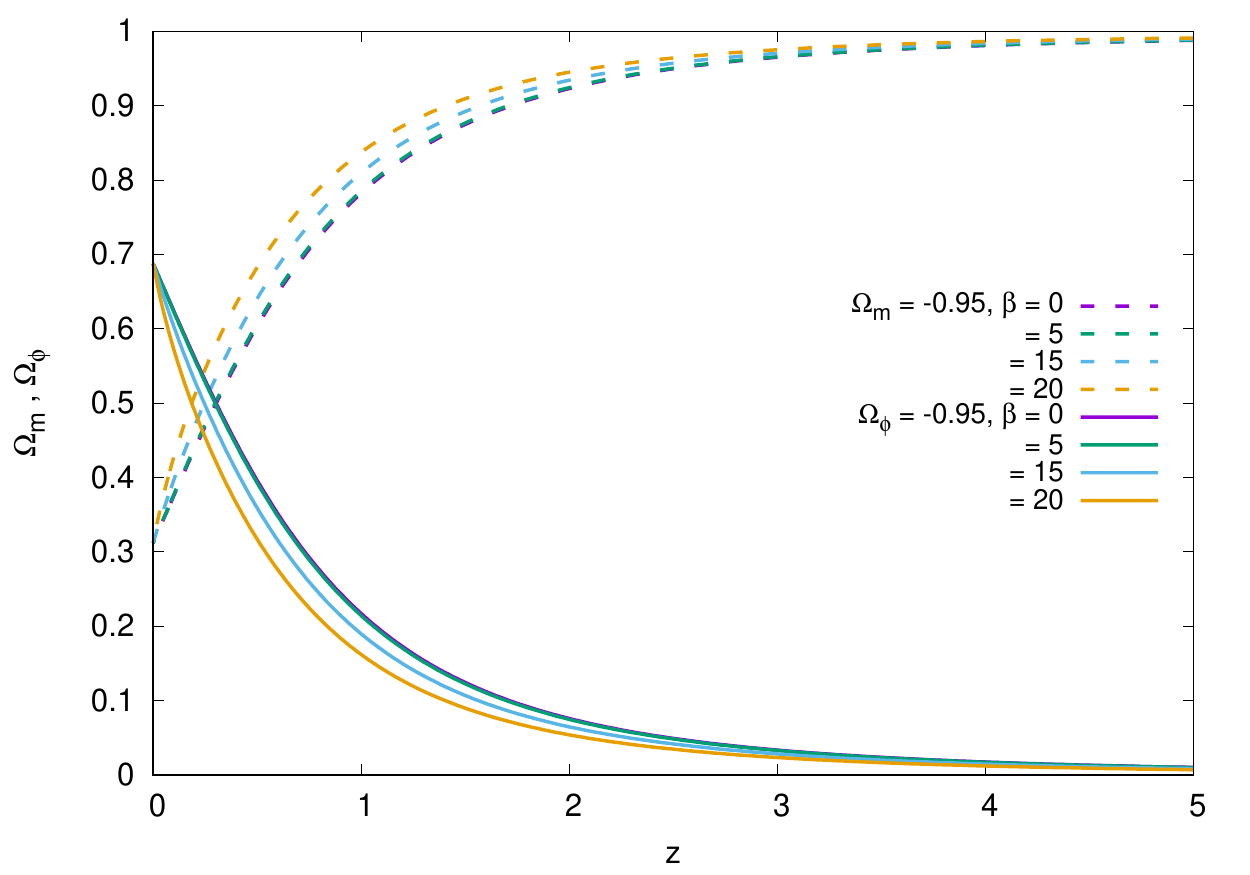}
\caption{\label{fig:density1}Plot of $\Omega_m$ and $\Omega_{\phi}$ for different values of $\beta$ parameters for the interaction A. All the $\alpha$ parameters are set to unity ($\alpha_0 = \alpha_1 = \alpha_2 = 1$) and the present value of the EOS of scalar field is chosen to be $w_{\phi} = - 0.95$.}
\end{figure}

\begin{figure}[] 
\centering
\includegraphics[width= \columnwidth]{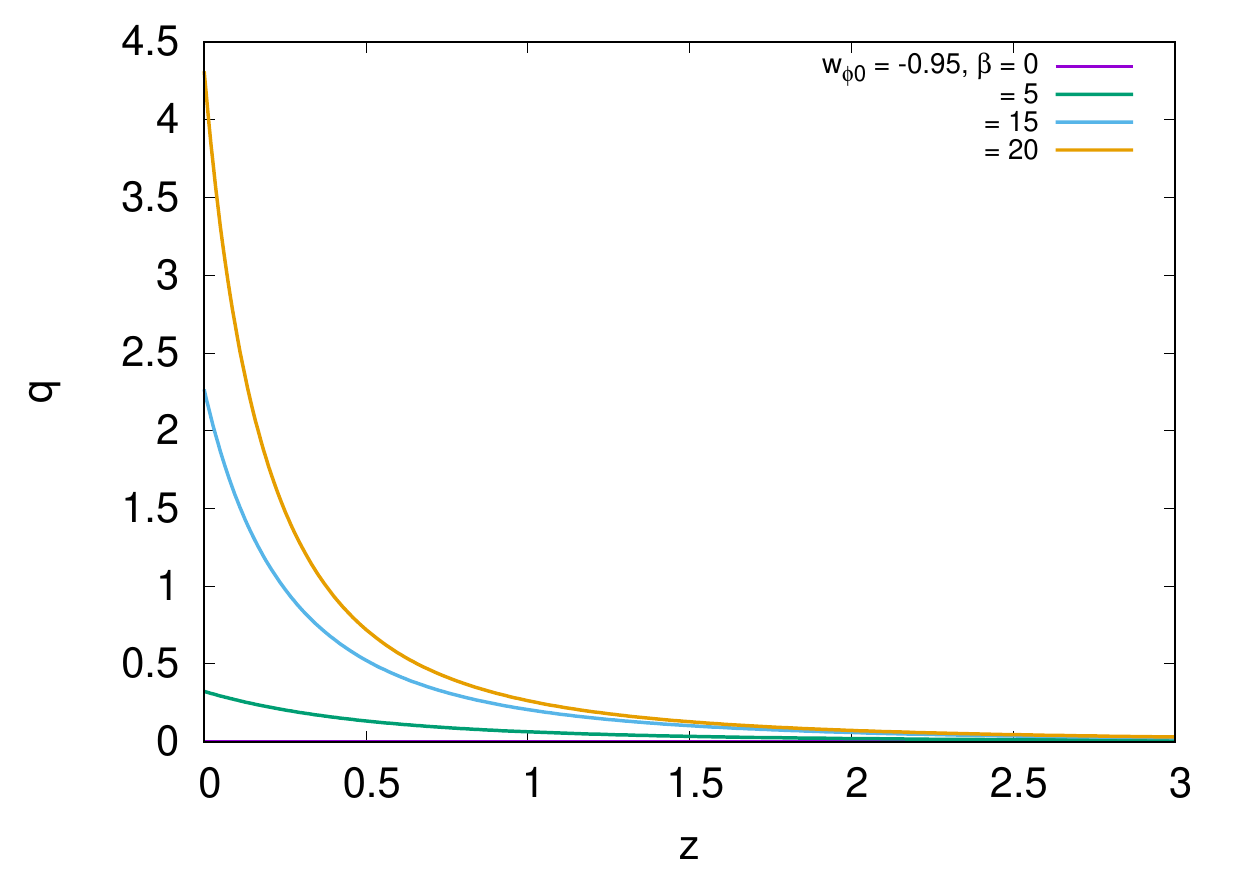}
\caption{\label{fig:q1}Plot of interaction variable $q$ for different values of $\beta$ parameters  for the interaction A. All the $\alpha$ parameters are set to unity ($\alpha_0 = \alpha_1 = \alpha_2 = 1$) and the present value of the EOS of scalar field is chosen to be $w_{\phi} = - 0.95$.}
\end{figure}

In the Fig.\ref{fig:density1} we have plotted the matter density parameter $\Omega_m$ and the scalar field density parameter $\Omega_{\phi}$ with respect to the redshift ($z$) for different values of $\beta$ parameter. It is interesting to note from Fig.\ref{fig:density1} that the matter and dark energy equality redshift decreases with increase of $\beta$. From this observation in the plot we expect to have a constrain on the allowed higher value of the parameter $\beta$ while we use the observations to constrain the cosmological parameters. Fig.\ref{fig:q1} shows the evolution of the interaction parameter $q$ for different values of $\beta$ parameter. In the remote past the interaction term $q$ was close  to zero for any value of $\beta$ but it has started to evolve recently. This nature of the interaction parameter suggests us that if there is transfer of energy from the dark matter to dark energy it has started very recently and this could be the reason why the universe at present is dominated by the dark energy.

\begin{figure*}[] 
\centering
\includegraphics[width= \textwidth]{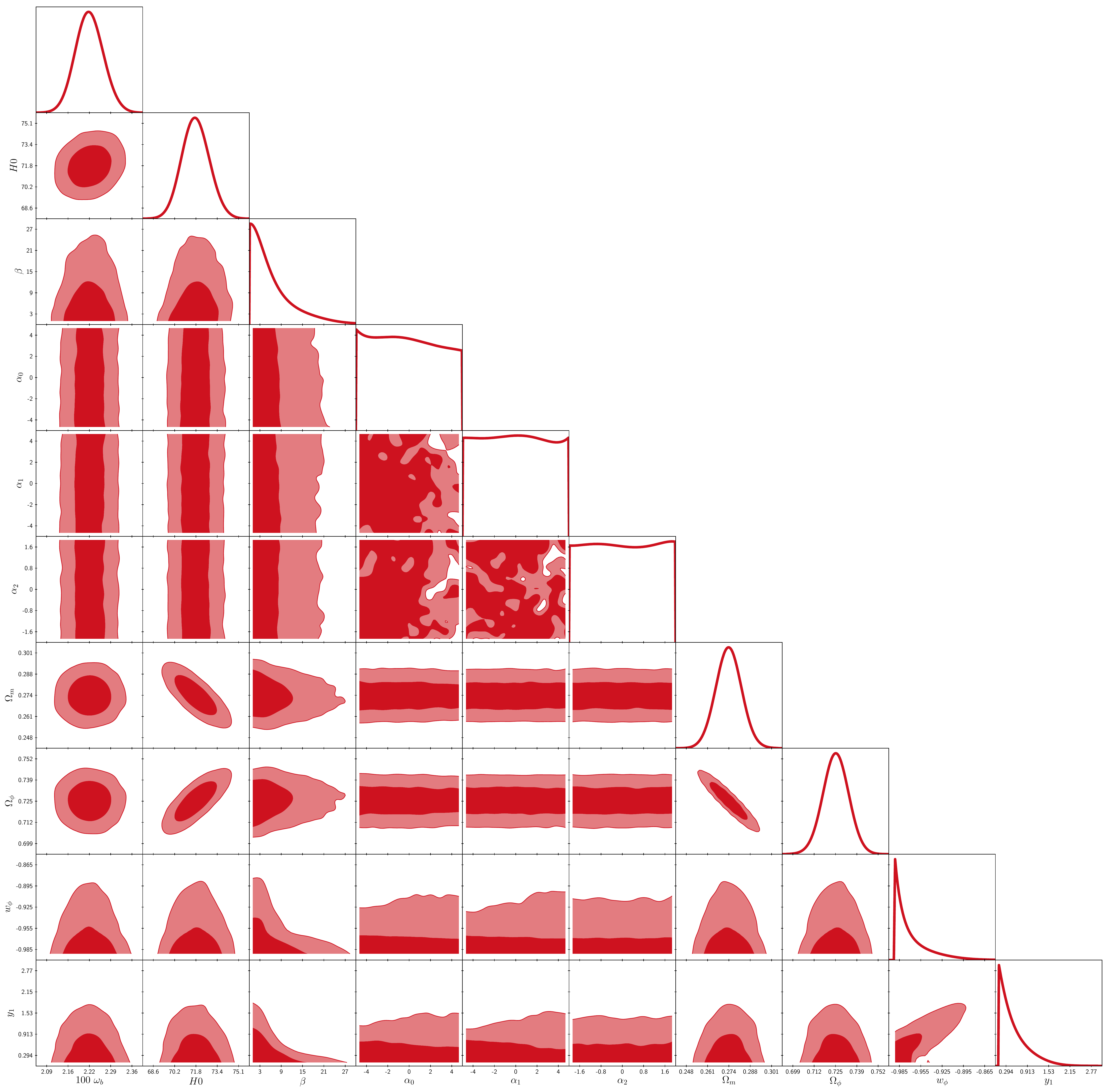}
\caption{\label{fig:triangle1}Plot of  Posterior (1D and 2D) distributions of the constrained cosmological parameters for interaction A. The datasets used are BAO+JLA+H(z) and a PLANCK15 prior is imposed on $\omega_b$ and $\omega_{cdm}$.}
\end{figure*}

The constrain on the cosmological parameters for the interaction A is shown in Fig.\ref{fig:triangle1}. We get back the same result of the \cite{roy2018new} on the constrain on $\alpha$ parameters. It is found that the cosmological parameters like, $\omega_b, \omega_{cdm}, H_0, \Omega_{\phi}, \Omega_m $ are very well constrained whereas the $\alpha$ parameters are unconstrained for these data sets. Table II is  the best fit values of the cosmological parameter for the interaction A. The interaction parameter $\beta$ is having a maximum cutoff value $0 \leq \beta \leq 7.62 $. 

\begin{table}[h]
\caption{Best fit value of the cosmological parameter for the Interaction A.}
\begin{ruledtabular}
\begin{tabular}{|c|c|}
Cosmological Parameters & JLA + BAO + $H(z)$ \\
\hline
$w_b$ & $2.23^{+0.0467} _{-0.467}$  \\
\hline
$H_0$ & $71.8^{+1.04} _{-1.09}$  \\
\hline
$\beta$ & $6.49^{+1.13} _{-6.49}$ \\
\hline
$\Omega_m$ & $0.274^{+0.00835} _{-0.00869}$ \\
\hline
$\Omega_{\phi}$ & $0.726^{+0.00869} _{0.00835}$ \\
$w_{\phi}$ & $-0.979^{+0.00357} _{-0.0209}$ \\
\hline 
$y_1$ & $0.444^{+0.1} _{-0.46}$

\end{tabular}
\end{ruledtabular}
\end{table}

\subsection{$Q = \beta (3 H^2 - \kappa \rho_{\phi}) \dot{\phi}^2 / H $}

\begin{figure}[h!] 
\centering
\includegraphics[width= \columnwidth]{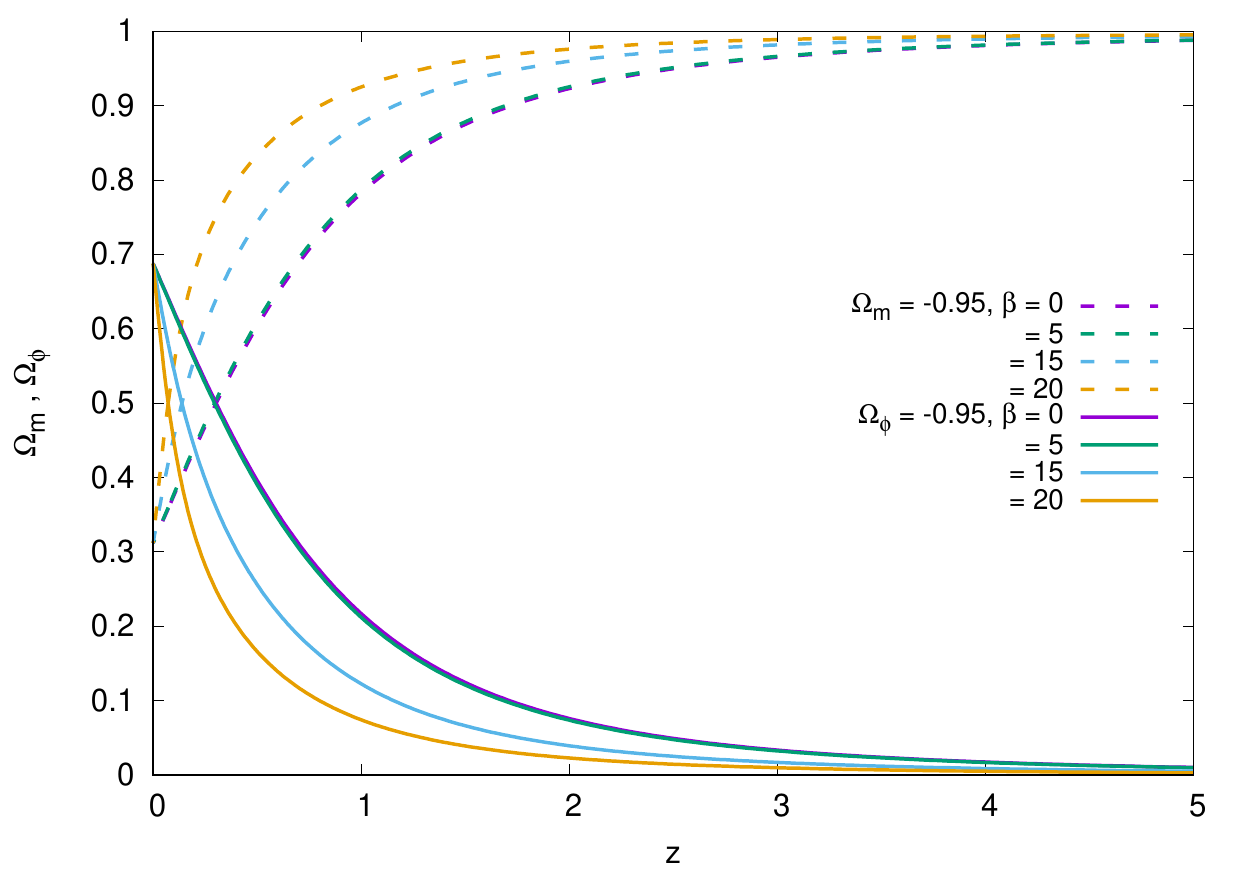}
\caption{\label{fig:density2}Plot of $\Omega_m$ and $\Omega_{\phi}$ for different values of $\beta$ parameters for the interaction B. All the $\alpha$ parameters are set to unity ($\alpha_0 = \alpha_1 = \alpha_2 = 1$) and the present value of the EOS of scalar field is chosen to be $w_{\phi} = - 0.95$.}
\end{figure}

\begin{figure}[h!] 
\centering
\includegraphics[width= \columnwidth]{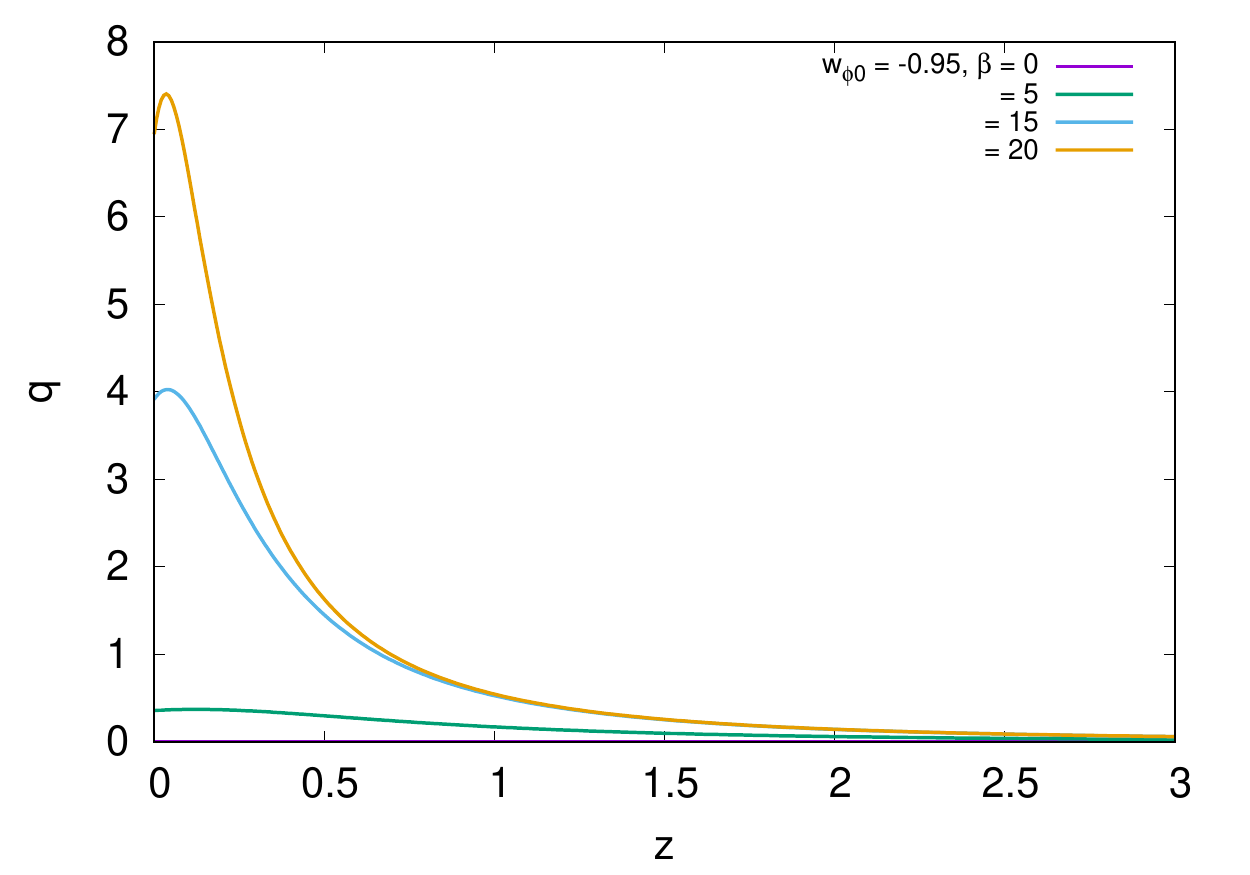}
\caption{\label{fig:q2}Plot of interaction variable $q$ for different values of $\beta$ parameters  for the interaction B. All the $\alpha$ parameters are set to unity ($\alpha_0 = \alpha_1 = \alpha_2 = 1$) and the present value of the EOS of scalar field is chosen to be $w_{\phi} = - 0.95$.}
\end{figure}

The plot of the $\Omega_m$ and $\Omega_\phi$ is shown in the figure Fig.\ref{fig:density2} for the interaction B with different values of the $\beta$
parameter. Like the interaction A similar nature of the  plot is observed. The dark matter and dark energy equality redshift decreases with increase in $\beta$ and hence a similar constrain on the maximum allowed value of the $\beta$ parameter is expected from the cosmological data analysis. In the Fig.\ref{fig:q2} the corresponding cosmological parameters for the interaction B is plotted. A interaction between the dark matter and the dark energy has started recently and a transfer of energy from the dark matter to dark energy happens due to this.

\begin{figure*}[h!] 
\centering
\includegraphics[width=\textwidth]{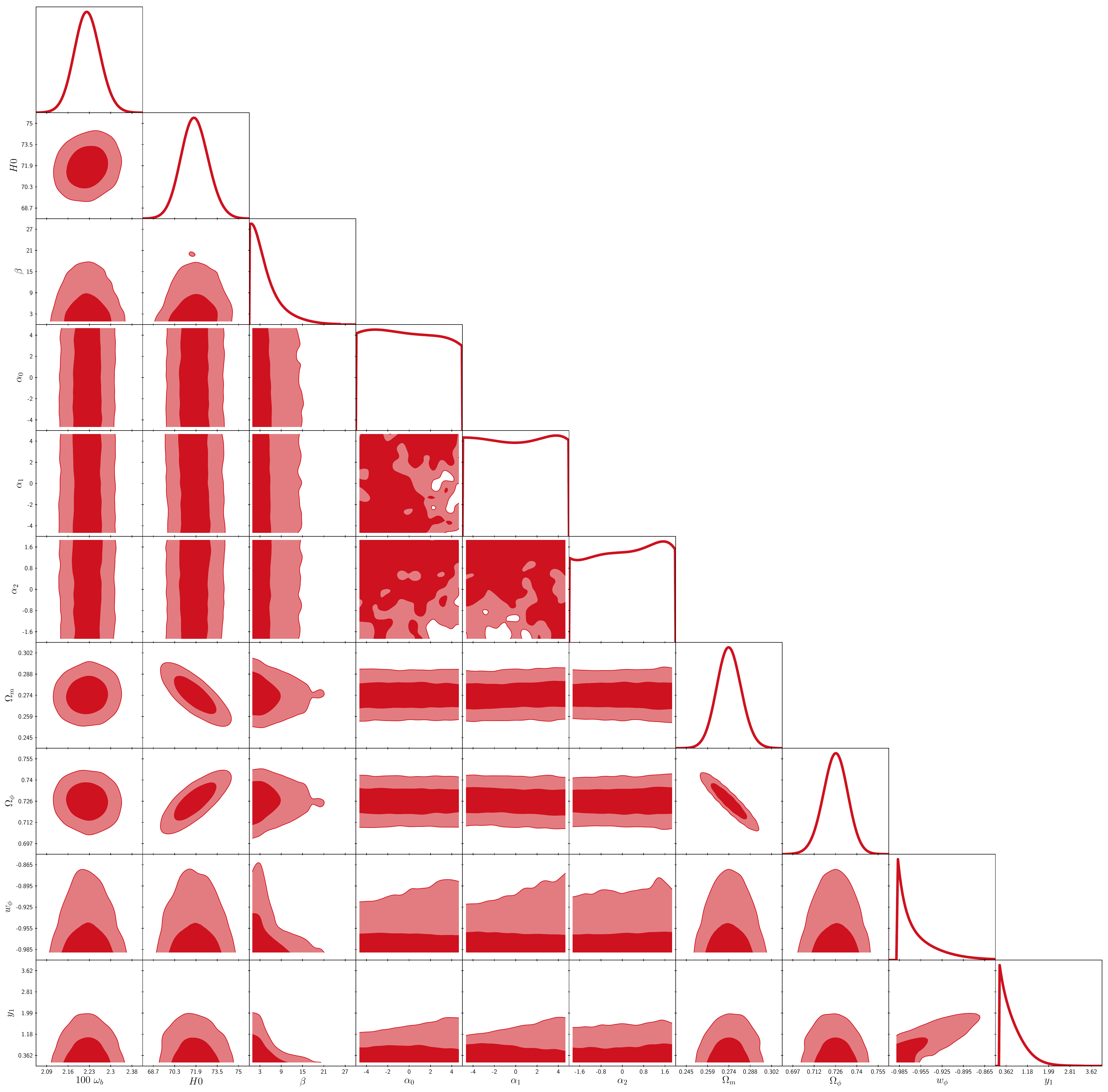}
\caption{\label{fig:triangle2} Plot of  Posterior (1D and 2D) distributions of the constrained cosmological parameters for interaction II. The data sets used are BAO+JLA + H(z) and a PLANCK15 prior is imposed on $\omega_b$ and $\omega_{cdm}$.}
\end{figure*}

Figure (\ref{fig:triangle2}) shows the constrain on the cosmological parameter for the interaction II. Similar to the Interaction B the $\alpha$ parameters are not constrained so the choice of potential remains arbitrary. This plot also indicates about a upper bound on the interaction parameter $0 \leq \beta \leq 5.481 $ hence a transfer of energy from dark matter to dark energy can not be arbitrary large. The best fit values are given in Table III. 

\begin{table}[h!]
\caption{Best fit value of the cosmological parameter for the Interaction B.}
\begin{ruledtabular}
\begin{tabular}{|c|c|}
Cosmological Parameters & JLA + BAO + $H(z)$ \\
\hline
$w_b$ & $2.23^{+0.0471} _{-0.485}$  \\
\hline
$H_0$ & $71.8^{+1.03} _{-1.1}$  \\
\hline
$\beta$ & $4.68^{+0.801} _{-1.1}$ \\
\hline
$\Omega_m$ & $0.274^{+0.0085} _{-0.0088}$ \\
\hline
$\Omega_{\phi}$ & $0.726^{+0.0088} _{0.0085}$ \\
$w_{\phi}$ & $-0.976^{+0.0051} _{-0.024}$ \\
\hline 
$y_1$ & $0.497^{+0.124} _{-0.543}$
\end{tabular}
\end{ruledtabular}\end{table}

\section{Conclusion} In a very recent work \cite{roy2018new},  it has been shown that the present cosmological observations are not enough to constrain any particular form of the quintessence potentials. The arbitrary nature of the quintessence potentials remains unresolved though we have entered an era of high precision cosmology. To search the favourable form of quintessence potentials in \cite{roy2018new} a general form of the quintessence potentials has been proposed. This general parameterization consists of three arbitrary parameters which are called as $\alpha $ parameters or the dynamical parameters as these parameters effect the cosmological dynamics. Different combinations of these $\alpha$ parameters corresponds to different potentials. The original idea was to constrains these $\alpha$ parameters and which will allow some one to find which class of the potentials are more favorable. But the present observations has failed to do so as it can not put any constrain on the alpha parameters.  

In this work, we have extended the above mentioned study of quintessence scalar field using the general parameterization to the interacting quintessence models.  We have tried to check if an interaction between the dark sectors can improve the scenario and chose a particular class of the potentials. Two different forms of the interaction are considered as for example since the functional form of the interaction is also arbitrary. While choosing the form of the interaction the particular importance was given to those functional form of the interactions which make the system equations simpler as our main aim is to check if a interaction term can effect the findings in \cite{roy2018new}. As far our knowledge we expect the result we obtained from this exercise will remain qualitatively same for any other interactions.

Our results reconfirms the findings in \cite{roy2018new} as our analysis also fails to constrain the $\alpha$ parameters. We have considered a positive prior on the interaction parameter $\beta$ as we assume the energy transfer happens from dark matter to dark energy. It is interesting to note that we have found an constrain on the upper bound on the $\beta $ parameter which tells us that the transfer of energy from DM to DE can not be arbitrarily large. This result is quite expected as in the Fig.\ref{fig:q1} one can notice that the matter and dark energy equality redshift decrease with increase in $\beta$ for both the cases of the interactions. The matte and dark energy equality can not be arbitrarily small so an upper bound on the $\beta$ is natural to find.

The morale of this exercise is that it not possible to break the degeneracy in the quintessence potentials even we consider an interaction between the dark sectors. This degeneracy can not be broken as the recent cosmological observations can only constrain the present value of the equation of state parameter (EOS). For a given sets of initial values there will be always a set of $\alpha$ parameters and $\beta$ parameter which will satisfy the cosmological observations. Unless there is cosmological data on the evolution of the EOS of the dark energy we will not be possible to constrain the form of the quintessence potentials.

\section{Acknowledgements} The work of KB was supported in part by the JSPS KAKENHI Grant Number JP 25800136 and Competitive Research Funds for Fukushima University Faculty (18RI009).



\bibliography{quint}
\clearpage
\end{document}